\documentclass{article}

\usepackage{fancyhdr}
\usepackage{amsfonts, amsmath}
\usepackage{hyperref}
\usepackage{graphicx}
\usepackage{color}
\usepackage{ctable}
\usepackage{wrapfig}
\usepackage{natbib}

\usepackage{amssymb}

\usepackage[nodots]{numcompress}

\bibpunct{(}{)}{;}{a}{,}{,}

\newcommand{\rr}[1]{\textrm{#1}}

\begin{document}
\title{Bias-Variance and Breadth-Depth Tradeoffs in Respondent-Driven Sampling}

\author{Sergiy Nesterko$^{\rm a}$$^{\ast}$\thanks{$^\ast$Corresponding %
author. Email: nesterko@fas.harvard.edu \vspace{6pt}} \hspace{0.1cm} and Joseph %
Blitzstein$^{\rm b}$\\\vspace{6pt}  $^{\rm a,b}$\em{Department of 
Statistics, Harvard University,}\\\vspace{6pt} \em{1 Oxford St., Cambridge, MA 02138, USA}
\\\vspace{6pt} }

\maketitle

\begin{abstract}
	Respondent-driven sampling (RDS) is a link-tracing network sampling 
	strategy for collecting data from hard-to-reach populations, such as	
	injection drug users or individuals at high risk of being infected 
	with HIV. The mechanism is to find initial participants (seeds), 
	and give each of them a fixed number of coupons allowing them to 
	recruit people they know from the population of interest, with a mutual
	financial incentive. The new participants are given coupons again 
	and the process repeats. Currently, the standard RDS estimator used in 
	practice is known as the Volz-Heckathorn 
	(VH) estimator. It relies on strong 
	assumptions about the underlying social network and the RDS process. 
	Via simulation, we study the relative performance of the plain mean and 
	VH estimator when assumptions of the latter are not satisfied, under 
	different network types (including homophily and rich-get-richer 
	networks), participant referral patterns, and varying number of 
	coupons. The analysis demonstrates that the plain mean outperforms the VH 
	estimator in many but not all of the simulated settings, including 
	homophily networks. Also, we highlight the 
	implications of multiple recruitment and varying referral patterns on the 
	depth of RDS process. We develop interactive visualizations of 
	the findings and RDS process to further build insight into the 
	various factors contributing to the performance of current RDS
	estimation techniques.
\end{abstract}

\section{Introduction}
It is often vital in current public health research to obtain accurate 
estimates of prevalences and other population averages in hard-to-reach 
populations \citep{lansky_developing_2007, frost_respondent-driven_2006}.
Such a
population is difficult to sample from using traditional sampling techniques 
such as simple random sampling or stratified sampling. A few examples of 
populations that may 
be hard-to-reach in some societies are: individuals at high risk of HIV
 \citep{abramovitz_using_2009}, 
injection drug users \citep{frost_respondent-driven_2006}, and men who have sex 
with men \citep{he_accessing_2008}. The target estimand could 
be, e.g., the proportion of diabetics among the HIV infected individuals, or the 
mean income of injection drug users in a certain area.

Respondent-Driven Sampling (RDS) is a technique for surveying hard-to-reach 
populations \citep{magnani_review_2005}, employing a link-tracing network 
sampling strategy to collect data from respondents belonging to the target population 
\citep{heckathorn0}.  A chain is formed when study participants 
recruit their acquaintances within the target population, by giving each participant a certain number of \emph{coupons} (with a financial incentive for both recruiter and recruitee) that can be used to recruit the next ``wave" of respondents. 

The process begins with the selection of a certain number of seeds (taken to be a convenience sample), and continues until either the process dies out, the study reaches the desired sample size, or the 
budget of the study is exhausted. Use of this method is rapidly increasing worldwide, with over 123 RDS-based studies performed in the 2003-07 time frame \citep{malekinejad_using_2008}. However, 
RDS poses serious data analysis challenges, especially because in most social networks there is \emph{homophily} (the tendency of people who are similar to form ties). In RDS, the data come from chains meandering 
through an underlying social network, and thus may be correlated in a very complex way, entangled with the  network structure. 

It is crucial to distinguish between RDS as a sampling scheme, and the estimators currently used along with RDS. Currently, the standard  estimator used with RDS is the \emph{Volz-Heckathorn} (VH) estimator, which was derived under a long list of strong assumptions, with the goal of obtaining an asymptotic unbiased estimator \citep{heckathorn1, heckathorn2, volzheck}. The asymptotic unbiasedness 
has been shown to be sensitive to assumption violations \citep{handcock}; here though our focus is on the bias and variance of RDS estimators for a fixed sample size, over a wide variety of network topologies, recruitment preferences, and numbers of coupons. 

%



Our work focuses on evaluating the relative performance of the conventional 
mean and the VH estimator under RDS settings, when 
the assumptions of the latter do not hold. We aim to provide intuition into how RDS behaves on different types of networks with different recruitment preferences, and to build broader understanding of the interplay between network structure, sampling processes, and estimation. 

Since there are so many assumptions to consider, possible network structures and  recruitment preferences, and random networks and RDS runs needed, summarizing the myriad of simulation results is complicated. To help make sense of this, we introduce several new visualization methods. Other studies also compare the performance of the plain mean and VH \citep{goelsalganik, handcock}. Our work builds on and differs from these in many ways, detailed later. In brief, our simulation setup allows flexibility in studying different recruitment structures by introducing a social space to use when generating 
networks and samples, creating dependence between the
network structure and the surveyed quantity. We also study the effects of varying numbers of coupons and uncertain degrees.

We especially emphasize violations of uniform recruitment, exact degrees being known, and dependence between network structure and the quantity of interest; these have not been treated in depth in the literature previously. The results can be viewed as an exploration of bias-variance tradeoffs (e.g., the VH estimator often reduces bias but at a heavy cost in terms of variance) and breadth-depth tradeoffs (obtaining more contacts per participant by increasing the number of coupons, vs.~obtaining longer chains through the network).

The simulated RDS processes match several 
empirically observed characteristics of actual RDS-based studies  
\citep{malekinejad_using_2008}.  As a further tool for understanding RDS under various settings, we have 
developed and made available online two interactive visualizations. One 
contains summarized findings from all performed simulations, and the other 
shows RDS process dynamics on networks 
of different types as a function of the surveyed quantity. Section 
\ref{sec:discussion} presents the relevant links.

The remainder of this work is organized as follows. In Section 
\ref{sec:background}, we describe 
the background of the problem in more detail, and review other 
recent studies. In Section \ref{sec:findings}, 
we summarize the findings and technical details of the simulation study, 
followed by detailed description of the study results. We group the findings 
according to the following features: homophily, rich-get-richer 
network topologies, number of coupons, sample size as a 
fraction of the total population, seed selection mechanism of the RDS process, 
and exact or stochastic degree reporting. Section \ref{sec:summarized} 
presents some overall findings, and Section \ref{sec:mechanics} delves further into the interpretations and visualizations of our findings.
Finally, Section \ref{sec:discussion} discusses some overall conclusions and limitations of the work.

\section{Background} \label{sec:background}

\subsection{The VH estimator}

The main idea behind the VH estimator \citep{volzheck} is to treat RDS as a 
random walk on an undirected network. It is well known from Markov chain theory that the stationary (equilibrium) probability of a node is then proportional to its degree. However, with multiple recruitment (i.e.,
multiple coupons per participant), the process is more like a branching 
process than a random walk. The VH 
estimator also assumes that the process converges quickly to stationarity (in practice it is very unlikely that a substantial ``burn-in"
time would be feasible, aside from the unsavory sound of ``burn-in" here). The VH estimator uses inverse probability weighting, as in a Hansen-Hurwitz estimator:
\begin{align}
	\hat{\theta}_{VH} = \left( \sum_{i \in I} D_i^{-1}\right)^{-1} \cdot \sum_{i \in I} D_i^{-1} X_i, \label{eq:heck}
\end{align}
where $I$ is the index set of all nodes in the sample, $D_i$ and $X_i$ are
 self-reported degree and quantity of interest of node $i$, respectively.  
This estimator was developed assuming single recruitment but is often
used without amendment in multiple recruitment settings  (with unclear justification). This estimator is based on the following assumptions:

\noindent  1. Respondents recruit uniformly from their 
    contacts in the population of interest. \label{itm:unifrecruit} \\
 2. The self-reported degrees $D_i$ are equal to the true degrees.  \label{itm:exactdegree} \\
  3.  There is single 
    recruitment (one coupon per participant). \\
  4. The chain may self-intersect (i.e., sampling is with replacement). \\
    5. Stationarity is quickly achieved (to wash away the initial bias from seed selection).

There are also several crucial assumptions about the underlying social network 
of acquaintances within the hidden population of interest. The network 
needs to be connected (so that everyone has the possibility of getting into the sample; for example, note that a ``hermit" of degree 0 could never be recruited, unless chosen as a seed). Moreover, network ties need to be reciprocal (as the social network is viewed as an undirected graph). In real implementations of RDS, however, the above assumptions are rarely satisfied:

\noindent 1. It is very unlikely that all 
    acquaintances of a respondent will get the same treatment when he or she 
    thinks about who to refer, e.g., recruiting a close friend may be more likely. Indeed, it may not even be easy to define precisely the set of possible people that a participant may recruit, since it's difficult to give a precise definition of ``acquaintance".  \\
2. Respondents' degrees are unlikely to be known accurately, even if the degrees are precisely defined. In practice, the personal network sizes must be estimated; this is often done by asking the respondent how many people with specific names that he or she knows within the population 
    of interest, and then scaling up  \citep{zheng_how_2006, mccormick_how_2010}. \\ 
  3. Typically, multiple recruitment is in place, often with 3 or more coupons. \\
  4. Most studies do not allow the same person to be recruited more than once. Even if self-intersection is allowed, participants may feel it is a waste of time to be resurveyed. 
  

\subsection{Previous studies}

The performance of the VH estimator in various settings is also evaluated in other studies \citep{goelsalganik, handcock}. We briefly review these and other related studies, in comparison with this work. 

The paper \citep{goelsalganik} evaluates the relative performance on simulated RDS processes in some specific networks. The authors show that even though the plain mean estimator generally exhibits more bias than the VH 
estimator, it performs better in terms of variance. Moreover, 
the variance of the VH estimator is often dramatically underestimated with 
the current technique \citep{salganik_variance_2006}, 
with resulting confidence intervals whose coverage rates are far below the 
nominal 95\% rates. These results are for a specific set of interesting 
networks, under 
VH-favorable conditions. In contrast, we study the performance of the VH 
estimator for a large ensemble of random networks of different types, under 
conditions where the VH assumptions hold to varying degrees. Our results 
confirm the large variance of the VH estimator, and show how the 
bias-variance tradeoff varies across settings.

The paper \citep{handcock} evaluates the VH estimator as 
a point estimator, assuming a  binary response.
The authors consider homophily, 
different seed selection and participant referral patterns as a function 
of underlying surveyed binary response, and discuss their effects on the bias 
of the VH estimator. They also examine the bias effect of RDS samples 
constituting a large fraction of total population (larger than 50\%). We focus on sample fractions less than 50\% since these seem  more likely 
in practice \citep{malekinejad_using_2008}. We also use a very different 
network generation method, designed so that we can investigate more 
classes of quantity dependent network topologies and conduct a sensitivity 
analysis for every network topology class. Additionally, we consider the effect of uncertain degrees, and perform breadth-depth analysis by simulating RDS  with varying numbers
of coupons per person. This enables new insights across a wide range of regimes.

In addition to the studies discussed above, there has been other work in this 
area \citep{lu_sensitivity_2010, amber_tomas_effect_2010}. The former considers simulating RDS  processes on a known network, but under VH-adverse conditions 
for RDS samples. The authors observe that the VH estimator is biased if 
recruitment preferences are determined by a covariate that is correlated 
with the surveyed quantity. We confirm this observation in our work, 
providing new ways to understand this phenomenon. The latter gives simulations similar to some of those discussed above,  but also includes several 
predecessors to the VH estimator.  Here, we focus on  the 
plain mean and the current VH estimator, but go into much greater detail 
into the reasons behind the differences in performance, and study MSE 
rather than just bias.


\section{Findings and simulation setup} \label{sec:findings}

\subsection{Summary of findings}

\textbf{Homophily.} For  homophily networks (where people who are similar in the quantity of interest are more likely to be connected), the plain mean uniformly outperforms the VH estimator, \emph{both} in terms of bias and variance, under all 
combinations of all the other features. This finding is a strong indication 
that dependence of referral structure and/or network topology on the quantity 
surveyed may severely undermine the performance of the VH estimator, making 
the plain mean a more appropriate estimator to use. We discuss the details of these findings in Section \ref{sec:homophily}, and discuss the 
possible reasons for the observed performance loss of the VH estimator in 
Section \ref{sec:mechanics}.

\textbf{Rich-get-richer.} For rich-get-richer networks (where people with a 
higher value of the quantity of interest tend to have more connections), the VH estimator outperforms the plain mean under all settings, 
except under inverse preferential referral (a typically unrealistic 
setting where respondents try to recruit people far from them in the social space). The plain mean has better variance 
properties under a uniform RDS recruitment pattern, but this is 
less consistently observable for smaller sampling fractions. Thus, for rich-get-richer networks, unless  inverse 
preferential recruitment holds, the VH estimator may be better than 
the plain mean. Detailed analysis and some  intuitions are 
given in Section \ref{sec:richgetricher}.

\textbf{Number of coupons.} We find that the number of coupons does not, in general, have a strong effect on the performance of the VH or plain mean estimator. We also study the effect of the number of coupons on the number of waves; see 
Section \ref{sec:numofcoupons}.

\textbf{Sampling fraction.} With sample size fixed at 300, close to the 
average sample size  in \citep{malekinejad_using_2008},  we  consider 
sampling fractions of  33\% and 10\%. Our findings suggest that with 
the considered absolute sample size, sampling fraction has little effect on 
observed relative performance of the plain mean and the VH estimators. We 
discuss the issues of absolute and relative sample size further in Section 
\ref{sec:percsamsize}.

\textbf{Seed selection.} As a general result, we observe that given other 
factors, the influence of seed selection on the relative performance of the VH 
and plain mean estimator is small. This is surprising, as proportional to 
degree seed selection should improve the performance of the VH estimator as 
it puts the process into the state when network nodes are sampled 
proportionally to degree. We conclude that seed selection is not a major 
factor determining the performance of the VH estimator, relative to other VH 
estimator assumption violations as defined by our study. 
The effects of seed selection are described further in Section \ref{sec:seed}.

\textbf{Exact and stochastic degree reporting.} Our simulations show that 
under a Poisson assumption, randomness in degree reporting as defined in our 
study does not substantially influence the performance of the VH estimator, 
given the other features. However, overdispersed degree reporting could 
 harm the VH estimator, while the plain mean estimator is not subject to 
degree information. Further investigation is necessary to fully evaluate the effects of error 
in degree reporting on the performance of the VH estimator. We describe some 
results for randomness in degree reporting in Section 
\ref{sec:degrep}.

\subsection{Details of simulation setup}

The mechanism of the simulations is as follows. For every network topology 
and sensitivity constant, we generate 500 networks. For 
each network and every RDS process feature value combination, we perform 
100 length 300 RDS simulations, recording the quantity 
of interest on each node. The quantities of interest are simulated as an i.i.d. 
sample from a Normal distribution with mean 175 and variance 100. For each such RDS simulation, 
an estimate of the population mean is calculated using the plain vanilla mean and 
VH estimators, and we use these result to study the bias and variance of the estimators. 

Simulations are performed in various settings, summarized in Table \ref{tab:simulfeat}. The features are: network topology, 
network size, RDS number of coupons, referral function, respondent degree 
reporting and RDS seed selection. 
\begin{table}
  \begin{center}
    \caption{Simulation features and their possible values.\vspace{2mm} } 
    \label{tab:simulfeat}
    \begin{tabular*}{1\columnwidth}{@{}l@{\extracolsep{\fill}}l@{}}
      \hline
      \hline
      Feature & Values \\
      \hline
      Topology & homophily, inverse homophily, rich-get-richer \\
      Network size & $\{1000, 3000 \}$ \\
      Number of coupons & $\{2, 3, \ldots, 6\}$ \\
      Referral function & preferential, inverse preferential, uniform \\
      Degree reporting & exact, stochastic\\
      Seed selection & uniform, proportional to degree \\
      \hline
    \end{tabular*}
  \end{center}
\end{table}
In all our simulations, we aim to generate networks with topology dependent on 
the quantity measured, for two reasons. First, the VH estimator implicitly assumes
that the degree is independent of the quantity 
measured,  so it is  appropriate to check the performance  when this is violated. Second, such scenarios are 
plausible in real life. For instance, people may tend to make friends with similar
people  (leading to homophily networks), or opposites may 
attract (leading to inverse homophily networks), or people may try to become 
friends with people with a high value of the quantity of interest (leading to rich-get-richer 
networks). 


Before describing the particular functions used for the network 
generation, we define some notation: let $d$ be Euclidean distance and
for node $i$ with quantity measured $x_i$, let $\chi$ be the set of all 
measurements in a given network, 
$\mathop{\rr{rank}}_{\chi}(x_i) = \left| \{ x_j \in \chi: \mbox{ } 
x_j \leq x_i\}\right|$, and $l_{ij}=1$ indicate the presence of an edge 
between vertices $i$ and $j$.  In our simulations, 
$\chi$ has size $\left|\chi\right|$ which is  1000 or 3000. We then 
create edges in a conditionally independent manner (given the $x_i$'s), with 
the following probabilities:
\begin{align*}
  \textrm{Homophily:}& \mbox{\hspace{0.3cm}} P(l_{ij} = 1) = \textnormal{invlogit} \left( -a \cdot d(x_i,x_{j})\right); \rr{ let } a \in (0.2, 1) \\
  \textrm{Rich-get-richer:}& \mbox{\hspace{0.3cm}} P(l_{ij} = 1) = \frac{b}{|\chi|} \max\left(\mathop{\rr{rank}}_{\chi}(x_i), \mathop{\rr{rank}}_{\chi}(x_j)\right); \rr{ let } b \in (0.1, 0.5) \\
  \textrm{Inv. homophily:}& \mbox{\hspace{0.3cm}} P(l_{ij} = 1) = -20 + \textnormal{invlogit}\left(c \cdot d(x_i,x_{j})\right); \rr{ let } c \in (0.8, 1.2)
\end{align*}
The ranges for the 
sensitivity parameters $a$, $b$ and $c$ are chosen so as to form networks that are 
connected, but not extremely dense. In our simulations, we set each 
sensitivity parameter to ten equidistant values from their respective ranges 
to study how the results change with varying network topology within its 
class. 

A key assumption of 
the VH estimator is that each respondent chooses  uniformly among his 
or her acquaintances. We 
explore this  using three respondent referral functions: preferential (participants tend to refer people close to 
them in the social space), inverse preferential (participants tend to 
refer people distant from them), and uniform (recruitment 
is uniform at random). We use the quantity measured to position individuals 
in the social space, and Euclidean distance as our metric. Let $X_i$ be the 
quantity recorded in the RDS sample on step $i$ and 
$N_i$ be the set of neighbors of node $i$, with all 
functions defined as zero for $x_{k}$ already in our sample, and for 
$k \notin N_i$. The particular 
functions used are:
\begin{align*}
  \textrm{Preferential:}& \mbox{\hspace{.1cm}} P(X_{i+1} = x_{k}|X_{i} = x_{i}) = d(x_i,x_{k})^{-c}\left/\sum_{j \in N_i}d(x_i,x_{j})^{-c} \right.; \rr{ let } c=1.5\\
  \textrm{Inverse pref.:}& \mbox{\hspace{.1cm}} P(X_{i+1} = x_{k}|X_{i} = x_{i}) = a^{d(x_i,x_{k})} \left/ \sum_{j \in N_i}a^{d(x_i,x_{j})} \right.; \rr{ let } a = e\\
	\textrm{Uniform:}& \mbox{\hspace{.1cm}} P(X_{i+1} = x_{k}|X_{i} = x_{i}) = 1 / | N_i |.
\end{align*}
Another assumption behind the VH estimator is that participants of the study 
accurately report how many people they know within the population of 
interest. This is unlikely to hold in reality, so we allow for uncertain degree reporting, by supposing that the degree of node $i$ is recorded as 
a draw from a Poisson($r_i$) distribution, where $r_i$ is the true degree.  Finally, we simulate the seed selection as either uniform across the network, 
or proportional to  degree. The latter rule seems more plausible in practice.

We perform simulations for networks of sizes 1000 and 3000, and vary the 
number of coupons from 2 to 6. The way we simulate recruitment 
is as follows. There is a 30\% probability that a respondent does not recruit 
anyone; if he or she does recruit, the number of recruitments is uniform over legal values, and whom to recruit is generated  based on the referral 
functions.

There are several pitfalls that must  be handled carefully. It is possible that the chain 
stops before reaching length 300. In this case, new seeds are selected and RDS is performed until the sample size reaches 300. This scenario is rare 
in our simulations due to the 8 seeds and good network connectivity. 
Another issue is that a node with degree 0 could be a seed. Such a seed cannot recruit anyone, but we give it degree 1 in  estimation to avoid dividing by 0.


\subsection{Homophily networks} \label{sec:homophily}

\begin{center}
	\begin{figure*}[h]
	\centering
  \includegraphics[width=1\textwidth]{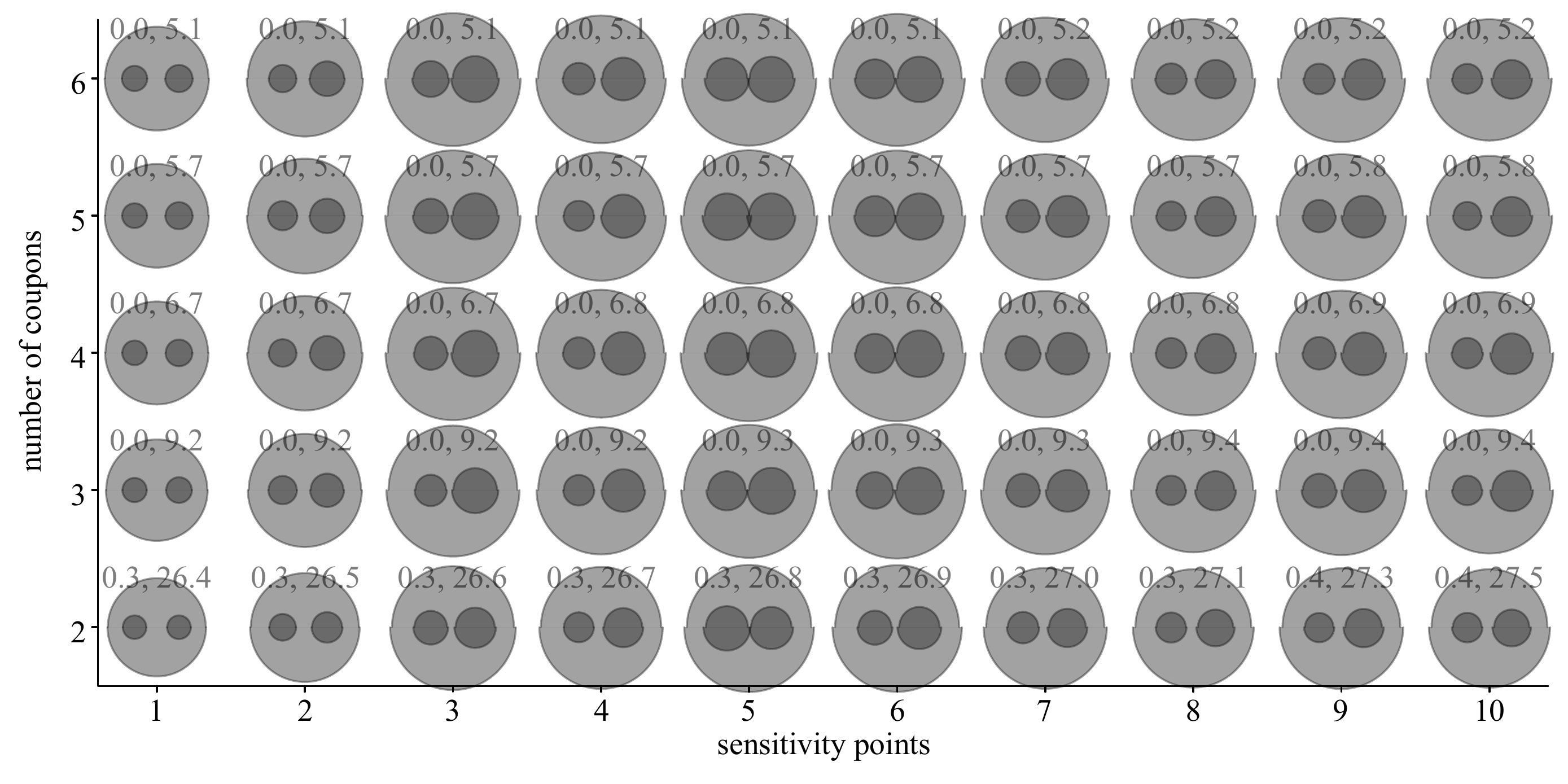}
	\vspace{-0.5cm}
	\caption{Relative performance of plain vanilla mean and VH estimators 
	as point estimators for homophily topology, network of size 1000, 
	uniform seed selection and preferential new respondent referral, 
	with RDS processes of 2 to 6 coupons, and network topology sensitivity 
	parameter at 10 different values. Every component of the plot consists 
	of upper and lower sets of 3 half circles. Large half circles 
	correspond to absolute difference in average MSE, left small half 
	circles correspond to absolute difference in squared bias, and right 
	small half circles correspond to absolute difference in variance. All 
	differences are between the VH and plain mean estimator. Grey 
	background color 
	means positive sign, and white  means negative sign. In this 
	plot, there are no instances of a white background. Upper sets of half 
	circles correspond to exact degree reporting, and lower sets correspond to random degree reporting. The pairs of numbers 
	separated by commas mean the average number of times the RDS process 
	had to restart, and the average number of waves needed to collect the 
	sample with particular network topology and number of coupons, 
	given other process features.}
	\label{fig-sens-1000-attr-unif-pref}
	\vspace{-0cm}
\end{figure*}
\end{center}

Figure \ref{fig-sens-1000-attr-unif-pref} gives a visualization of the relative 
performance of the plain vanilla mean and the VH estimator as point estimators for 
RDS processes with varying numbers of coupons, uniform seed selection and 
preferential participant referral mechanism, simulated on networks with 
homophily with sensitivity constant at ten different values. Note that 
network connectivity decreases as the sensitivity constant increases.

Strikingly,  the plain mean outperforms the VH estimator  across the board  in \emph{both} bias and variance. Some  reasons for this can be gleaned from Figure \ref{fig-deg-qua}. 
Better-connected nodes are in the center of the distribution of the 
underlying quantity, so  the RDS process visits there often. The VH 
estimator suffers performance loss due to occasional 
low degrees of nodes in the center of the distribution, while the plain mean does not even use degree information!


Here 500 networks have been simulated for 
every sensitivity constant index, for computational reasons. This is small relative to the very large number of possible networks of size 1000 with specified topology. 
Therefore, figures such as Figure \ref{fig-sens-1000-attr-unif-pref} should 
be construed as giving general trends in relative estimator performance.

\subsection{Rich-get-richer networks} \label{sec:richgetricher}
\begin{center}
	\begin{figure*}[h]
	\centering
	  \includegraphics[width=1\textwidth]{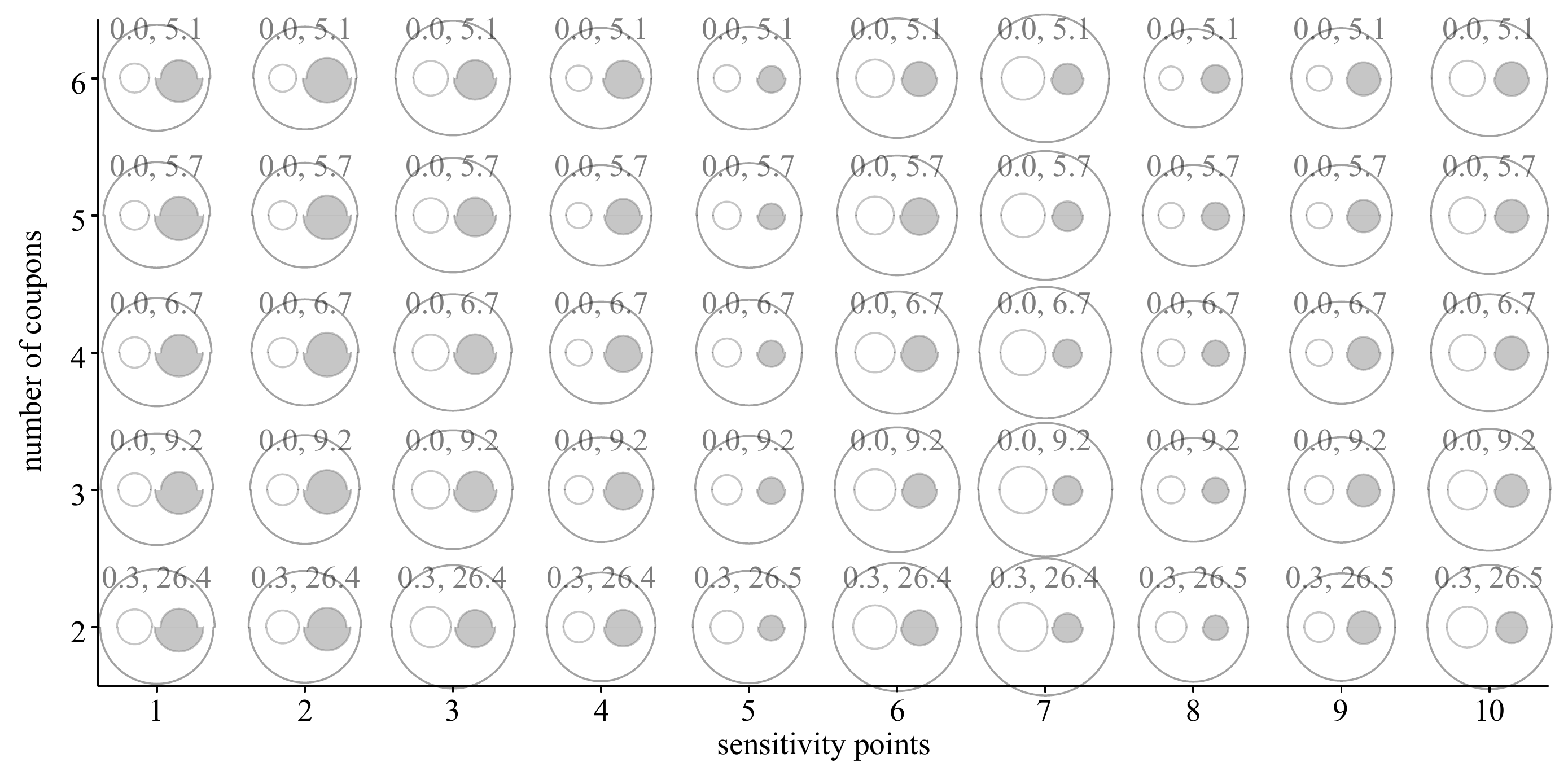}
	\vspace{-0.5cm}
	\caption{Relative performance of plain vanilla mean and VH estimators 
	as point estimators for rich-get-richer topology, network of size 
	1000, proportional to degree seed selection, uniform new respondent 
	referral, with RDS processes of 2 to 6 coupons, and network topology 
	sensitivity parameter at 10 different values. Graphical elements 
	and numbers are as described in Figure 
	\ref{fig-sens-1000-attr-unif-pref}.} \label{fig-sens-1000-power-degProp-unif}
\end{figure*}
\end{center}

Figure \ref{fig-sens-1000-power-degProp-unif} shows the relative performance 
of the plain mean and VH estimator for rich-get-richer networks of size 1000, 
and RDS processes with proportional to degree seed selection, and uniform 
participant referral. Note that the VH estimator outperforms the mean 
estimator across the board in MSE. The reason can be seen in Figure \ref{fig-deg-qua}. Nodes with high degrees correspond to the tail of the distribution  of the quantity of interest. Proportional to degree seed selection causes 
the process to visit the right tail  more frequently, and 
the VH estimator does a better job discounting the corresponding observations 
because it weights them by inverse degree. The resulting bias reduction is 
sufficient to compensate for a worse variance.
Figure \ref{fig-sens-1000-power-degProp-unif} demonstrates that the VH 
estimator has  lower bias across the board, and the plain mean has better 
variance, especially in the case of stochastic degree reporting. However, for 
this case, bias reduction of the VH estimator is significant enough to give 
it an edge against the plain mean estimator. Here, random degree reporting helps the plain mean estimator gain ground against 
the VH estimator, since this smoothes out the low weights given by the VH estimator to the 
tail measurements. 

The only setting when the VH estimator underperforms the plain mean for 
rich-get-richer networks is inverse preferential referral. In that setting, participants are referring people who are far away in the distribution, and thus the left tail of the distribution 
is visited more often (see Figure \ref{fig-deg-qua}). Then the
plain mean outperforms the VH estimator, as the VH estimator induces negative bias by up-weighting the 
left tail observations. The online resources described in Section \ref{sec:mechanics} provide interactive visualizations of this phenomenon. 

\subsection{Number of coupons} \label{sec:numofcoupons}

We find that the number of coupons does not, in general, have a strong effect on the performance of the VH or plain mean estimator. Also, under the standard recruitment 
scheme, the RDS process does not need to be restarted to reach the necessary sample size, except in the 2-coupon case. 
 The number of seeds has been set to agree with a survey of  RDS studies 
\citep{malekinejad_using_2008}. To collect 300 observations, 2-coupon RDS processes yield unrealistically large numbers of waves. For 3-coupon RDS processes, 9 waves are obtained on 
average, which concurs with the average reported in \citep{malekinejad_using_2008}. RDS processes with 4, 
5 and 6 coupons on average yield numbers of waves of 7, 6 and 5,
respectively.


%


\subsection{Sampling fraction} \label{sec:percsamsize}

Our study considers two possible values for the sampling fraction. The sample size 
is 300, and network sizes are either 1000 or 3000, so the sampling fractions  are 33\% and 10\%.
\begin{center}
	\begin{figure*}[h]
	\centering
	  \includegraphics[width=1\textwidth]{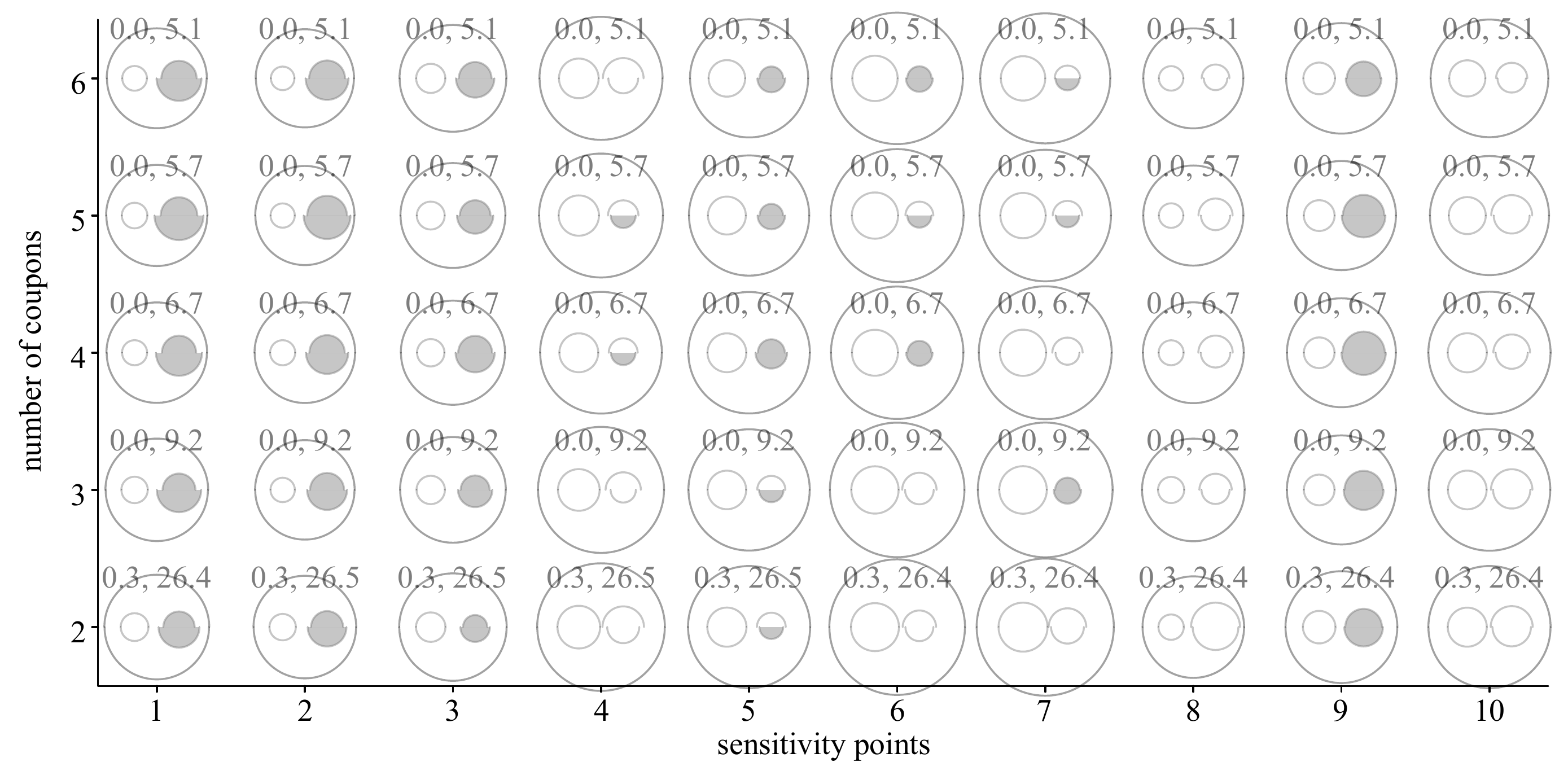}
	\vspace{-0.5cm}
	\caption{Relative performance of plain vanilla mean and VH estimators 
    as point estimators for rich-get-richer topology, network of size 
    3000, proportional to degree seed selection, uniform new respondent 
    referral, 
    with RDS processes of 2 to 6 coupons, and network topology 
    sensitivity parameter at 10 different values. Graphical elements 
    and numbers are as described in Figure 
    \ref{fig-sens-1000-attr-unif-pref}. The VH 
    estimator dominates the plain mean in all setups here.} \label{fig-sens-3000-power-degProp-unif}
\end{figure*}
\end{center}
Generally, with the considered absolute sample size, 
sampling fraction has no effect on the patterns of estimator relative 
performance for every combination of simulation features. However, 
Figure \ref{fig-sens-3000-power-degProp-unif} demonstrates that the VH 
estimator also has better variance than the plain mean in some instances,
while Figure \ref{fig-sens-1000-power-degProp-unif} corresponds to 
smaller network size and does not exhibit this observation.  These findings suggest that for the sample sizes currently collected via RDS, 
sampling fraction is not a deciding factor for the choice of the estimation 
technique. This is encouraging, as in practice the sampling fraction is typically unknown.
\subsection{Seed selection} \label{sec:seed}

In our simulations, we consider two seed selection mechanisms. The first is 
selecting the seeds at random from the set of network nodes, and the second 
is selecting the seeds with probability proportional to degree. Ideally, 
proportional to degree seed selection should favour the VH estimator, 
because it puts the process in ergodicity mode right away.

The idea behind exploring seed selection feature is to see how much of an 
improvement proportional to degree seed selection provides to the VH 
estimator in the settings when other assumptions it is based on are violated. 
We do not observe any cases when it does help the VH estimator outperform the 
plain mean. That is, at any combination of other simulation features, the 
picture of relative performance of the VH and plain mean estimators is the 
same regardless of seed selection mechanism. 

Overall, we find that the possible improvement of the proportional to degree 
seed selection is overshadowed by other factors. We have even found 
some indication (under inverse homophily topology) that proportional to 
degree seed selection is detrimental for the VH estimator. Our online visualization resources can be used to explore this further (see Section 
\ref{sec:mechanics}).

\subsection{Exact and stochastic degree reporting} \label{sec:degrep}
\begin{center}
  \begin{figure*}[h]
    \centering
    \includegraphics[width=1\textwidth]{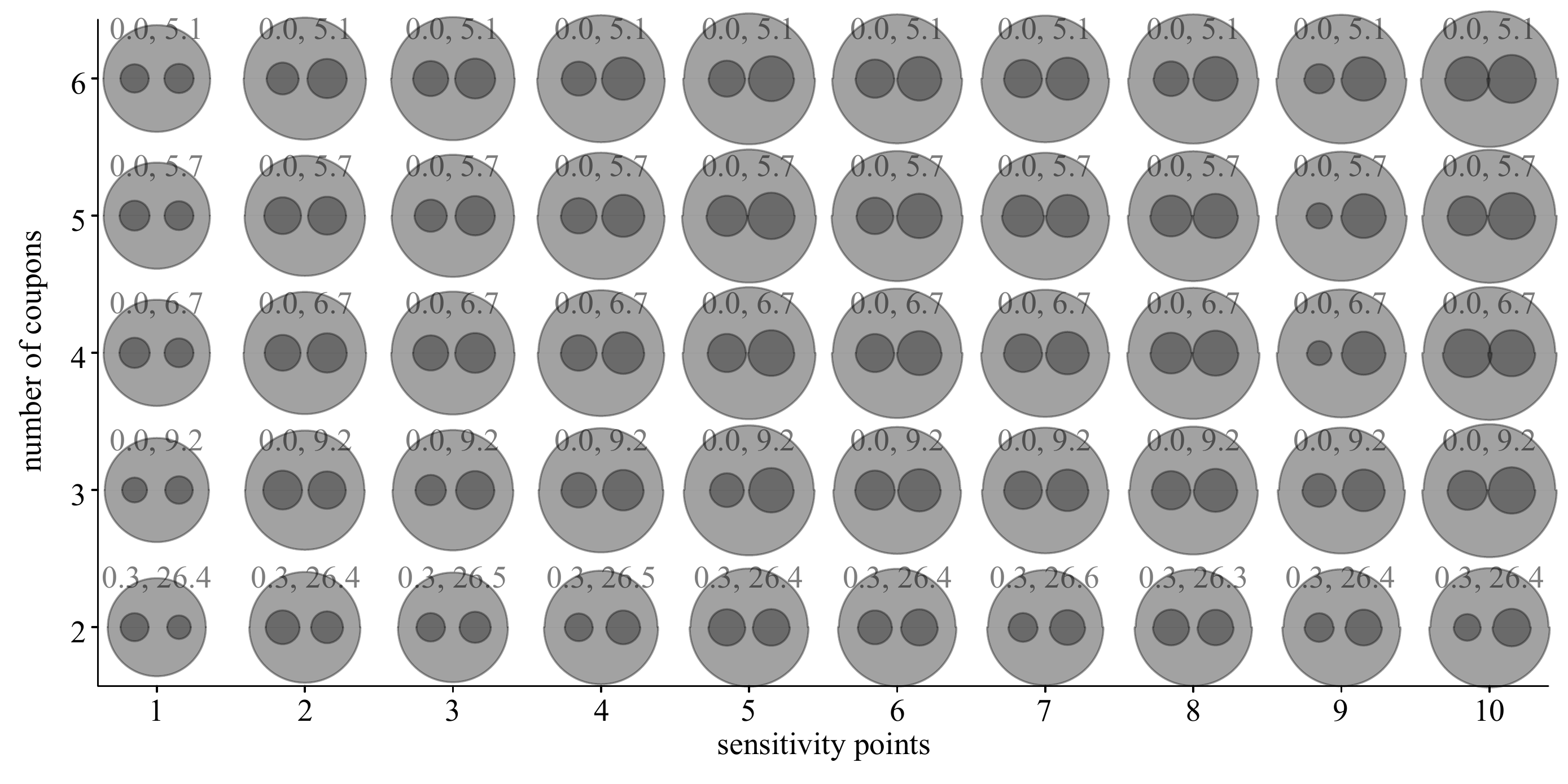}
    \vspace{-0.5cm}
    \caption{Relative performance of plain vanilla mean and VH estimators as 
    point estimators for homophily topology, network of size 3000, 
    proportional to degree seed selection, preferential new respondent 
    referral, with RDS processes of 2 to 6 coupons, and network topology 
    sensitivity parameter at 10 different values. Graphical elements and 
    numbers are as described in Figure \ref{fig-sens-1000-attr-unif-pref}. 
    Stochastic degree 
    reporting harms the performance of the VH estimator in this case.}
    \label{fig-sens-3000-attr-degProp-pref}
    \vspace{-0cm}
\end{figure*}
\end{center}

In our simulations, reported values are drawn from a Poisson distribution with mean equal to true degree.  With random degree reporting as defined in our study, there is 
little effect on the relative performance of the estimators.  Under homophily, stochastic degree reporting further diminishes the 
performance of the VH estimator against the plain mean estimator for both 
sampling fractions considered (see Figures \ref{fig-sens-1000-attr-unif-pref} 
and \ref{fig-sens-3000-attr-degProp-pref}), and it has almost no effect  
for inverse homophily and rich-get-richer topologies. Still, it is possible that the 
current techniques of degree estimation may carry overdispersed error 
\citep{zheng_how_2006,mccormick_how_2010}; the 
effect of uncertain degree reporting in  real RDS 
studies remains largely unknown. 

\section{Overall performance comparison} \label{sec:summarized}

The simulation results analyses presented so far have been carried out using  
sensitivity plots such as Figure 
\ref{fig-sens-1000-attr-unif-pref}. It is also useful to consider
relative performance averaged over the network topologies sensitivity 
constants. We present the relevant findings in terms of summary tables 
containing average MSE for the plain mean and VH estimator for all 
simulation feature value combinations, constructed by network size and 
number of coupons.

\begin{table}[h]
  \caption{Average MSE for plain mean (numbers to the left from semicolons) 
  and VH estimator (numbers to the right from semicolons; numbers to the 
  right from commas correspond to random degree reporting) for different 
  simulation configurations for networks of size 3000 and 3-coupon RDS. Cases when the VH estimator outperforms the plain mean are 
  displayed in italic. There is a large discrepancy for rich-get-richer network 
  topology and inverse preferential referral. Under homophily, the 
  VH estimator is dominated by the plain mean under all settings.} \label{tab:3000-3coupons}
  \begin{center}
    \begin{tabular}{r c c c}
      \hline \hline
      & \multicolumn{3}{c}{referral function} \\
      Network, seed & unif & pref & inv pref \\
      \hline
      homo, unif s & 14.46; 46.51, 48.57 & 28.73; 71.48, 73.54 & 3.08; 25.84, 29.28 \\
      homo, pr to deg s & 8.83; 21.52, 22.03 & 16.84; 30.22, 30.52 & 2.13; 15.98, 18.4 \\
      inv homo, unif s & \emph{1.05; 0.32, 0.32} & \emph{0.99; 0.28, 0.28} & 2.4; 3.32, 3.32 \\
      inv h, pr to deg s & \emph{1.07; 0.33, 0.33} & \emph{1.01; 0.3, 0.3} & 2.45; 3.84, 3.84 \\
      r-g-r, unif seed & \emph{4.08; 0.24, 0.25} & \emph{13.11; 8.62, 8.61} & 2.35; 43.75, 43.91 \\
      r-g-r, pr to deg s & \emph{4.24; 0.24, 0.24} & \emph{18.44; 12.11, 12.1} & 2.35; 43.6, 43.77
    \end{tabular}
  \end{center}
\end{table}

Table \ref{tab:3000-3coupons} summarizes the relative performance of the 
plain mean and VH estimator for networks of size 3000, and 3-coupon RDS. Under homophily, the VH estimator is 
dominated by the mean in all settings. For rich-get-richer networks with inverse 
preferential referral, the plain mean does much better than VH. Under all other scenarios,  VH  outperforms the plain mean. In these simulations, random degree reporting has little effect on the performance of the VH estimator except for a small detrimental effect in homophily networks.

\begin{table}[h]
  \caption{Average MSE for plain mean (numbers to the left from semicolons) 
  and VH estimator (numbers to the right from semicolons; numbers to the right 
  from commas correspond to random degree reporting) for different simulation 
  configurations for networks of size 1000 and 6-coupon RDS. Cases 
  when the VH estimator outperforms the plain mean are displayed in italic. 
  Overall, there is little change in relative performance of the 
  two considered estimators from what is observed in Table 
  \ref{tab:3000-3coupons}.} \label{tab:1000-6coupons}
  \begin{center}
    \begin{tabular}{r c c c}
      \hline \hline
      & \multicolumn{3}{c}{referral function} \\
      Network, seed & unif & pref & inv pref \\
      \hline
      homo, unif s & 12.88; 40.08, 43.52 & 18.55; 49.08, 52.55 & 6.11; 29.02, 33.12 \\
      homo, pr to deg s & 8.71; 20.77, 21.98 & 12.79; 25.33, 26.3 & 4.12; 16.55, 18.99 \\
      inv homo, unif s & \emph{2.72; 1.53, 1.51} & \emph{2.5; 1.33, 1.31} & 4.34; 8.75, 8.76 \\
      inv h, pr to deg s & \emph{2.98; 1.63, 1.61} & \emph{2.77; 1.4, 1.38} & 4.58; 9.87, 9.9 \\
      r-g-r, unif seed & \emph{3.2; 0.31, 0.33} & \emph{9.6; 5.01, 4.99} & 4.24; 24.29, 24.53 \\
      r-g-r, pr to deg s & \emph{3.29; 0.3, 0.31} & \emph{13.59; 7.31, 7.28} & 4.28; 23.83, 24.07
    \end{tabular}
  \end{center}
\end{table}

Table \ref{tab:1000-6coupons} contains summary of relative performance of the 
VH and plain mean estimator for networks of size 1000 and 6-coupon RDS. Similarly to  3-coupon RDS, the VH estimator underperforms on homophily networks under all scenarios. 

\section{Interpretation and visualization} \label{sec:mechanics}

Our simulation methods allow us easily to explore the bias-variance trade-off for the VH and plain mean estimator 
for a particular network topology sensitivity constant, and a given 
combination of other RDS features. For example, Figure \ref{fig-3000-attr-5-degProp-unif-cert} shows the squared bias, 
variance and MSE histograms of the plain mean and VH estimators for simulated 
RDS processes on networks with size 3000, network topology sensitivity constant 
5, proportional to degree seed selection and uniform participant referral.
\begin{center}
  \vspace{-0.5cm}
  \begin{figure*}[h]
    \centering
    \includegraphics[width=1\textwidth]{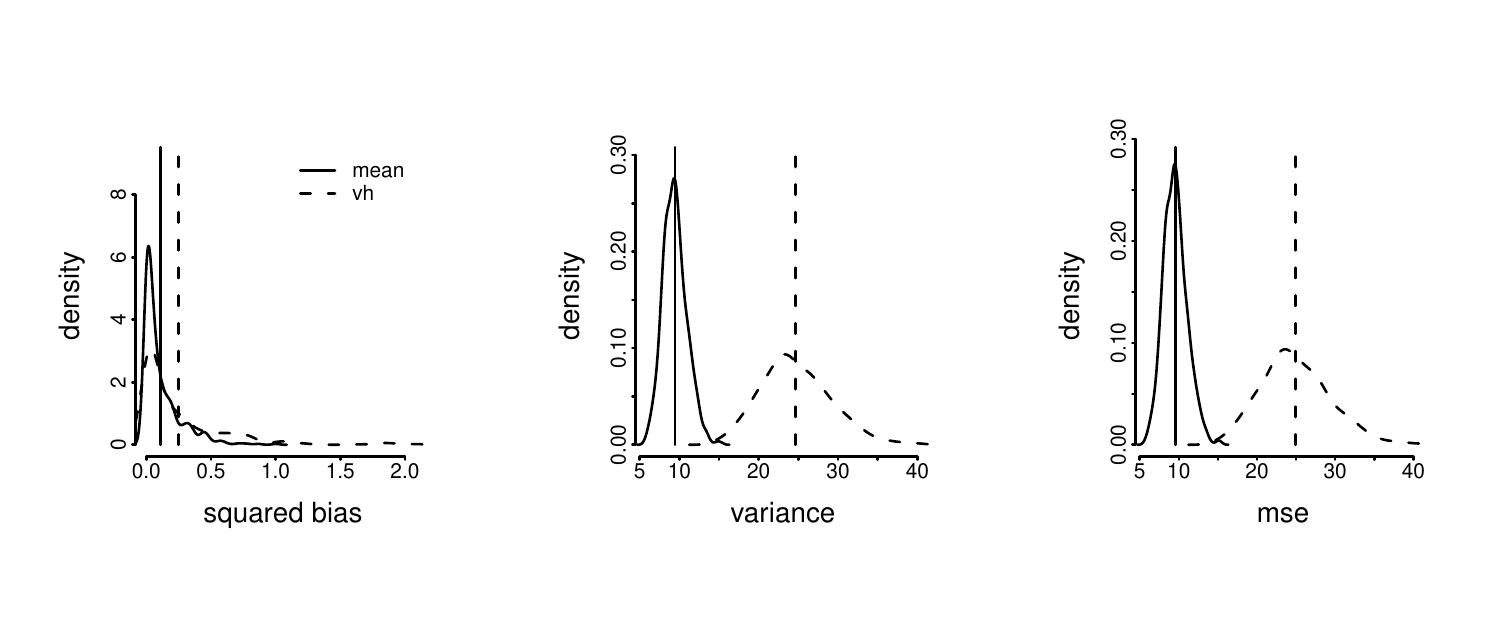}
    \vspace{-1cm}
    \caption{Estimated densities of squared biases, variances and MSE of 
    plain vanilla mean and VH estimators for RDS, homophily network of 
    size 3000, topology sensitivity constant 5, 3 coupon structure, 
    proportional to degree seed selection, uniform referral and exact 
    degree reporting. Straight lines represent density means. The mean 
    outperforms the VH estimator in all aspects, with a much 
    lower variance, and similar squared bias.}
    \label{fig-3000-attr-5-degProp-unif-cert}
    \vspace{-0cm}
\end{figure*}
\end{center}
Such figures help give insight into what performance components of an 
estimator suffer a larger loss relative to another on a given network topology and RDS features. 
 Figure \ref{fig-deg-qua} displays 
the relationship between quantity of interest and degrees of bearing vertices 
for different network topologies. 

\begin{center}
  \vspace{-1cm}
  \begin{figure*}[h]
    \centering
    \includegraphics[width=1\textwidth]{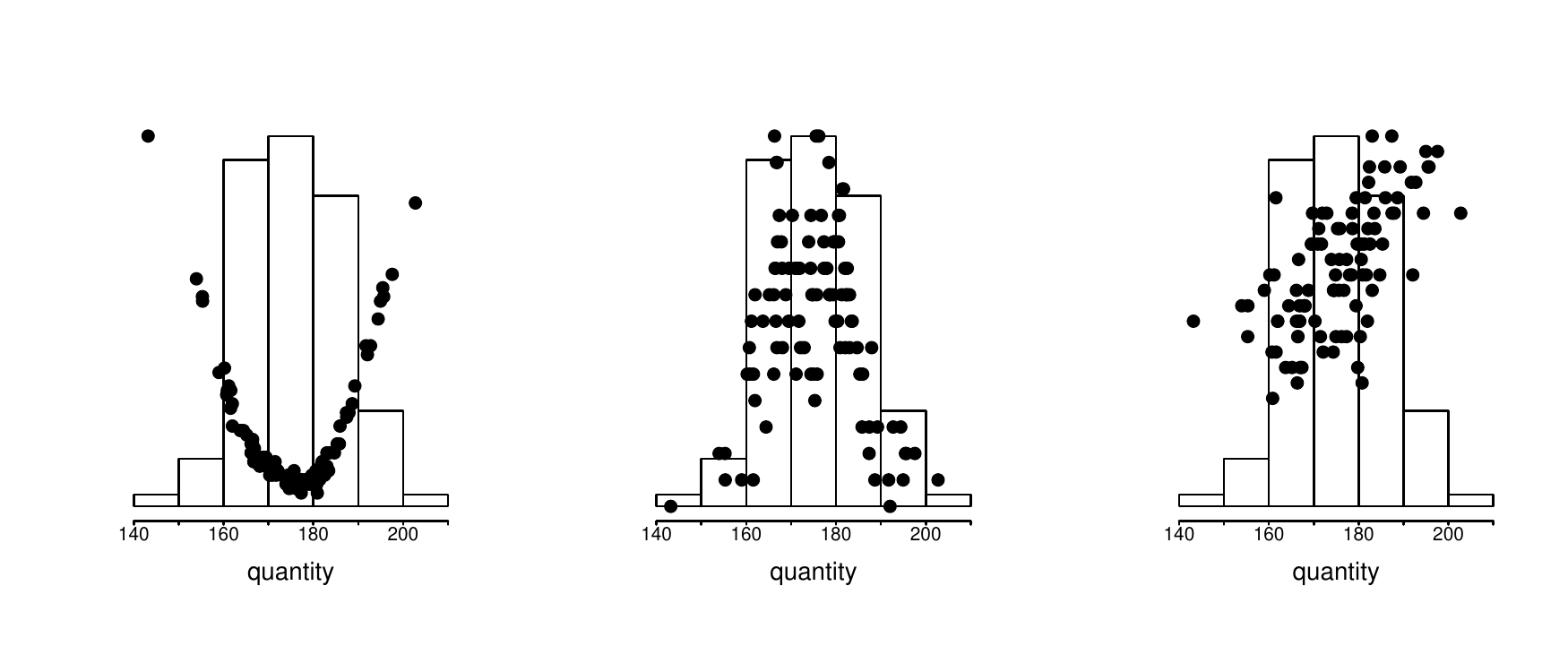}
    \vspace{-1.5cm}
    \caption{Histogram of RDS quantity of interest and scatterplot of 
    normalized degrees, for inverse homophily (left), 
    homophily (center) and rich-get-richer (right) networks.}
    \label{fig-deg-qua}
    \vspace{-.5cm}
\end{figure*}
\end{center}

To develop relevant intuition, let us consider an example of how Figure \ref{fig-deg-qua} is used. A notable finding from Table 
\ref{tab:1000-6coupons} is that the VH estimator outperforms the mean  when the network is explored with uniform or preferential referral 
function, on inverse homophily and rich-get-richer topologies. With uniform and 
preferential referral functions, nodes with high degrees get visited more 
often, so more tail measurements get into the RDS sample 
(see Figure \ref{fig-deg-qua}). The VH estimator weights observations by 
inverse degree, thus \emph{discounting} the tail measurements. The plain mean, on the other hand, gives 
equal weights to all observations in the sample. This makes the plain mean 
estimator have large bias and  larger variance than the VH 
estimator.

The VH estimator shines if the underlying network 
topology is such that vertices bearing tail measurements have high degrees. 
This is because it gives high weights to measurements from the middle of 
distribution, thus having both low bias and variance. This is well 
illustrated with Figure \ref{fig-3000-invAttr-5-degProp-unif-cert}, which shows 
inverse homophily. Under preferential and uniform referral 
functions, high degree vertices are visited more often and thus tail 
measurements get into the RDS sample more frequently 
(see Figure \ref{fig-deg-qua}). This hinders the performance of 
the plain mean and gives an edge to the VH estimator.

Analogously, for homophily networks, the plain mean outperforms VH
(see Figure \ref{fig-3000-attr-5-degProp-unif-cert}). In this setting, the 
mean wins by weighting all measurements equally as the RDS process stays 
near the center of distribution most of the time, while the VH estimator down-weights them.
\begin{center}
	\vspace{-0.5cm}
	\begin{figure}[h]
	\centering
	  \includegraphics[width=1\textwidth]{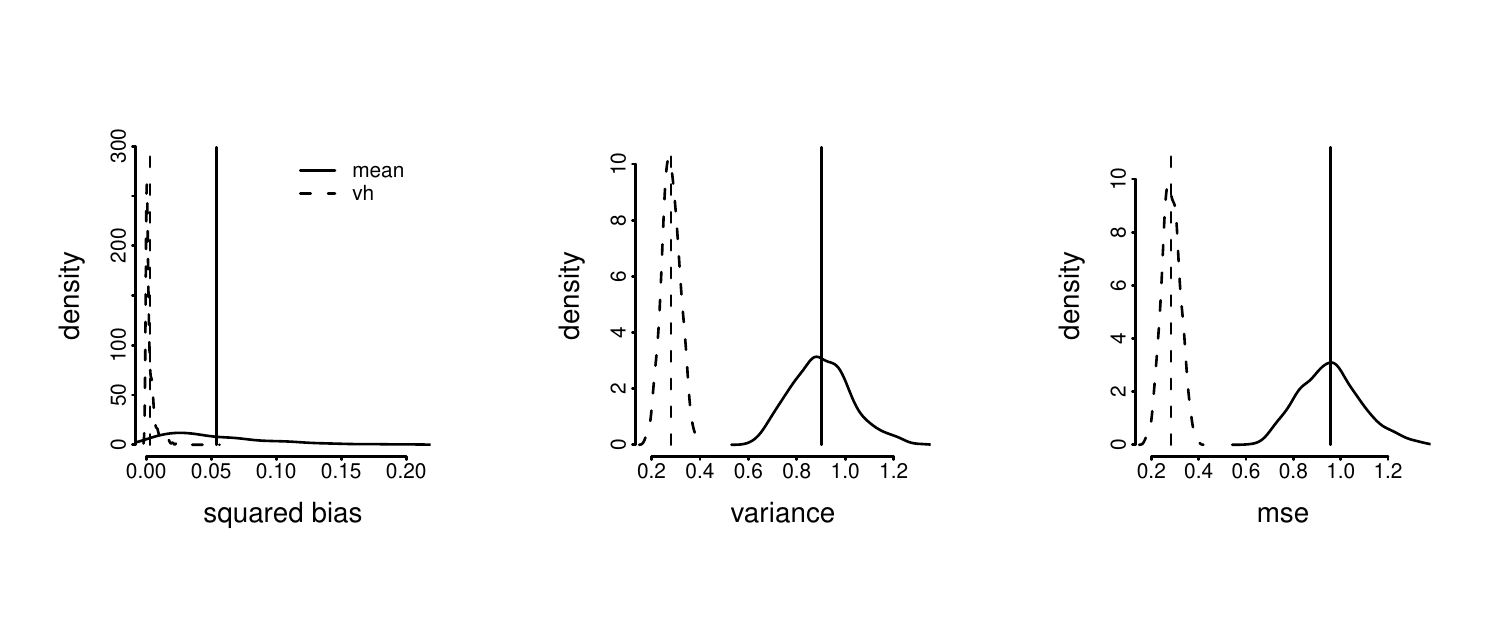}
	\vspace{-1.5cm}
	\caption{Estimated densities of squared biases, variances and MSE of 
	plain vanilla mean and VH estimators for RDS, inverse homophily 
	network of size 3000, topology sensitivity constant 5, 3 coupon 
	structure, proportional to degree seed selection, uniform referral 
	and exact degree reporting. Straight lines represent density means. 
	The VH estimator outperforms the plain mean in all 
	aspects, with a lower variance and squared bias.}
	\label{fig-3000-invAttr-5-degProp-unif-cert}
	\vspace{-0cm}
\end{figure}
\end{center}
Weighting observations by inverse degree puts the VH estimator in its most 
unfavorable situation when operating under inverse preferential referral 
function. When this function governs the RDS process, respondents try to 
refer next participants far from them in the distribution of the quantity of 
interest. The most vivid example is rich-get-richer networks with inverse 
preferential referral, when the mean outperforms the VH estimator by more than 
tenfold in terms of MSE (for example, see the bottom right of Table 
\ref{tab:3000-3coupons}). This happens because the left tail is frequently visited due to inverse preferential referral function 
(see Figure \ref{fig-deg-qua}), and the VH estimator overweights these 
observations because they have low degrees. This induces large negative bias, 
thus putting the VH estimator far behind.

We have made available online two supplementary visualizations: an interactive visualization of the plain mean and VH estimator tradeoffs for all settings described in this paper, with additional results presented for smaller scale simulations, and an interactive visualization of RDS processes with varying  network topologies,  as a function of the quantity 
    surveyed; see \href{http://incontemplation.com/go/rdsdynamic1}{http://incontemplation.com/go/rdsdynamic1} and \href{http://incontemplation.com/go/rdsdynamic2}{http://incontemplation.com/go/rdsdynamic2}.

\section{Discussion} \label{sec:discussion}

In this work, we have explored several key aspects of RDS for various network 
topologies dependent on the quantity surveyed. This makes it possible to compare estimation 
performance of various simulation 
features. We attempt to quantify the extent to which degree information is needed for adequate estimation, and identify the driving features that make one estimator underperform and another dominate in various settings. 

A limitation of our work is that it focuses on a univariate quantity of interest, with a small number of particular methods of generating network topology dependent on the quantity of interest. It would be interesting to develop  alternative methods with different network topologies but similar associations of  degree and  quantity of interest. 

Our findings suggest that using inverse-to-estimated-degree weightings for RDS estimation is questionable under many circumstances. In other simulations we have also explored compromise estimators, with the VH weights raised to a fractional power. These estimators smoothly interpolate between the plain mean and the VH estimator, but there is much scope for developing and selecting between various  compromise estimators. The results show that special care should be given to the role of homophily in RDS, and the interplay between the network structure, recruitment preference patterns, and the quantity of interest.

\section{Acknowledgements}
We thank Edoardo Airoldi, Victor DeGruttola, Richard Garfein, and Susan Little for helpful comments. Joseph Blitzstein acknowledges NIH P01 AI074621 for partial support, and Sergiy Nesterko acknowledges the Natural Sciences and Engineering Research Council of Canada via Postgraduate Scholarship 90350447 for partial support.

\bibliographystyle{plainnat}
\bibliography{bib-designbased}
\end{document}